\documentclass[a4paper,12pt]{amsart}

\usepackage{globaldef}

\usepackage[T1]{fontenc} 
\usepackage[utf8]{inputenc}

\usepackage{parskip}
\usepackage{listings}
\usepackage{ytableau}
\usepackage{cite}

\usepackage{mdframed} 

\hypersetup{ 
	pdftitle={integrable systems, programming and so on},
	pdfsubject={High Energy Physics}, 
	pdfauthor={thiago araujo},
	pdfkeywords={gauge; susy; strings; fields; cft; python},
	colorlinks=true,    
    linkcolor=myPurple, 
    citecolor=myPurple, 
    urlcolor =myPurple, 
    linktocpage=true    
}

\definecolor{myPurple}{RGB}{90, 74, 120}
\definecolor{myBlue}{RGB}{15, 75, 110}
\definecolor{myRed}{RGB}{191,97,106}
\definecolor{myDarkGray}{RGB}{216, 222, 233}
\definecolor{myLightGray}{RGB}{236, 239, 244}

\definecolor{c1}{RGB}{129, 162, 193} 
\definecolor{c2}{RGB}{216, 222, 233} 
\definecolor{c3}{RGB}{236, 239, 244} 
\definecolor{c4}{RGB}{59, 66, 82}
\definecolor{c5}{RGB}{76, 86, 106}

\newtheorem{proposition}{Proposition}
\newtheorem{conjecture}{Conjecture}
\newtheorem{lemma}{Lemma}

\DeclareMathOperator{\Res}{Res}
\DeclareMathOperator{\Tr}{Tr}

\DeclarePairedDelimiter{\bra}{\langle}{\rvert}
\DeclarePairedDelimiter{\ket}{\lvert}{\rangle}

\DeclarePairedDelimiterX{\bracket}[2]{\langle}{\rangle}{#1\vert#2}
\DeclarePairedDelimiterX{\bbracket}[2]{\langle\!\langle}{\rangle\!\rangle}{#1\vert#2}

\usepackage{slashed}                
\usepackage{simplewick}             
\usepackage{fontawesome}
\usepackage{tikz-cd}
\usepackage[textwidth=20mm]{todonotes}

\begin{document}


\title[Slavnov products, KP and BA functions]{More on Slavnov Products of Spin Chains and KP Hierarchy Tau Functions}

\author{Thiago Araujo}

\address{
Universidade Federal Fluminense, 
Instituto de Ciências Exatas, Departamento de Física
Volta Redonda, RJ, Brazil
}

\email{\texttt{\href{thiaraujo@id.uff.br}{thiaraujo@id.uff.br}}}

\begin{abstract}
 Connections between classical and quantum integrable systems are
 analyzed from the viewpoint of Slavnov products of Bethe states. It
 is well known that, modulo model dependent aspects, the functional
 structure of Slavnov products generally takes the form of
 determinants. Building on recent results on the structure of rational
 and trigonometric models, we show that, provided certain conditions
 are satisfied, the Slavnov product of a given model can be
 interpreted as a tau function of the KP hierarchy, thus extending
 known results in a more general setting. Moreover, we show that
 Slavnov products can be expanded in terms of other tau functions. We
 also prove that their homogeneous limit can be systematically
 expressed as a Wronskian of functions related to the eigenvalues of
 the transfer matrices.  Finally, we compute the Baker–Akhiezer
 functions associated with these Slavnov products and show that, apart
 from a universal multiplicative factor, they admit a closed
 determinantal representation.
\end{abstract}


\maketitle

\setcounter{tocdepth}{1}
\tableofcontents

\section{Introduction}

One important ingredient in the Algebraic Bethe Ansatz is the scalar
product of Bethe states. This object plays a central role in the
analysis of quantum integrable systems. Many important results have
been collected on this topic, see~\cite{Korepin:1993kvr} and
references therein. In this text, we focus on a powerful and elegant
aspect of these products: scalars products in the algebraic Bethe
ansatz can be expressed in terms of determinants.

More specifically, we investigate a very special type of correlation
function in quantum integrable spin chains, the so called
\emph{Slavnov product}~\cite{Slavnov1989}. These objects can be seen
as building blocks of more sophisticated correlation functions and are
instrumental in the study of norms of Bethe wave
functions~\cite{Korepin:1993kvr}. But more importantly for the
purposes of the present work, Slavnov products seem to be a bridge
connecting quantum and classical integrable system. This is the
problem we explore in this work.

Connections between classical and quantum integrable systems are a
long-standing research area, and have been addressed from a wide range
of perspectives~\cite{Wu:1975mw, Its:1992bj, korepin2000the, Foda2009,
  Alexandrov:2011aa}.  In~\cite{Foda:2009zz}, the authors started the
analysis of the connections between the Slavnov products of the
Heisenberg XXZ spin chain and the Kadomtsev–Petviashvili (KP)
classical integrable hierarchy. Their analysis has been extended in
several different directions and applications, for
example~\cite{Wheeler:2010vmq, Foda:2010, Takasaki:2010qm,
  Foda:2012wn, Foda:2012wf}.  In~\cite{Araujo:2021ghu,
  Araujo:2024klz}. The author of the current paper also investigated
some aspects of this research problem. In particular, the relation
between integrable hierarchies and quantum integrable systems such as
the Q-Boson integrable system and the Temperley-Lieb open spin chains.

The main ingredients for the Slavnov product are the Bethe
wavefunctions themselves. As such, the specific details of these
objects depend on the model, symmetries and their representations, and
on the boundary conditions. Nevertheless, the general
functional structure of Slavnov products across radically
different models is basically the same. More specifically, modulo some
multiplicative factors, all these expressions take the form of
determinants, and it is this basic property that allows us to
manipulate these objects and prove that they are KP and/or Toda tau
functions.

Belliard and Slavnov~\cite{Belliard:2019bfz} have thoroughly
investigated this property for spin chains with rational and
trigonometric \(R\)-matrices, and basically answered ``\emph{why
scalar products in the Algebraic Bethe Ansatz have determinant
representations}''. Using their results, we now want to understand
``\emph{why and how (some) scalar products in the Algebraic Bethe
Ansatz are KP and Toda tau functions}''. We prove that, as long as the
integrable system satisfies the conditions established
in~\cite{Belliard:2019bfz} -- along with some additional restrictions
-- the Slavnov product is guaranteed to satisfy the KP integrable
hierarchy. Additionally, we discuss some immediate consequences of
this result.

Let us now provide a brief overview of the content and structure of
our work. In Section 2, we offer a review of the main findings
presented in the work of Belliard and
Slavnov~\cite{Belliard:2019bfz}. In addition, we use this section to
carefully establish and clarify the notation that will be employed
throughout the remainder of this paper, as well as in related future
investigations.

Section 3 presents two new results and a conjecture. First, we rewrite
relevant formulas established in~\cite{Belliard:2019bfz} and prove
that these Slavnov products are KP tau functions expressed in terms of
alternant determinants. We also discuss some contrasts between this
formula and certain results that have previously appeared in matrix
theory investigations.  These tau functions are written in terms of
two sets of parameters. The first set consists of arbitrary complex
numbers, and the second set consists of the Bethe roots of the spin
chain. In Section 3, we also find a basis for the Slavnov products,
and this basis is itself formed by tau functions. The results of
Section 3 allow us to investigate the case where the complex
parameters are close to the Bethe roots.

In this section, we also conjecture (and only present some general
evidence) that these tau functions indicate that we are, in fact,
dealing with a multicomponent KP hierarchy, and that the linear
equations discussed by Belliard and Slavnov~\cite{Belliard:2019bfz}
describe a reduction of this larger integrable hierarchy.

Section 4 discusses the homogeneous limit of the Slavnov product.  In
this case, all complex parameters condense to a single value, and we
show that Slavnov products have a Wronskian expression written in
terms of functions related to the eigenvalues of the transfer matrix.
Section 5 presents some explicit examples for small spin chains, and
we see that even in those cases the expressions for the tau functions
become overwhelming very quickly.

Finally, in Section 6, we use the Japanese formula to investigate the
Baker–Akhiezer function associated with these tau functions. Using
known integral formulas for the tau functions, we write them in terms
of Miwa coordinates. The most important result in this section is the
explicit expressions for the Baker–Akhiezer functions. We discuss some
interesting consequences and open problems in Section 7.

\section{Scalar products as determinants}

To ensure the paper is as self-contained as possible, this section
reviews the main arguments of the work of Belliard and
Slavnov~\cite{Belliard:2019bfz}. Let us start with a set of arbitrary
complex parameters \(\bm{u} = \{u_j\}_{j=1}^{n+1}\), and define
\(n+1\) sets \(\bm{u}_j=\bm{u}\setminus \{u_j\}\). In the Algebraic
Bethe Ansatz context~\cite{Korepin:1993kvr, Slavnov:2019aba}, we can
now define \(n+1\) off-shell Bethe vectors \(\ket{\Psi(\bm{u}_j)}\),
i.e., Bethe states where the algebraic Bethe equations are not imposed
on the parameters \(\bm{u}_j\). Additionally, we must also consider a
set of Bethe roots \(\bm{v} = \{ v_k \}_{k=1}^n\), that is, parameters
\(\bm{v}\) that satisfy the algebraic Bethe equations. Finally, we
define the on-shell Bethe vectors \(\ket{\Psi(\bm{v})}\).

\subsection{Determinant representation}
The action of the transfer matrix \(\mathcal{T}(z)\), that is an
Hermitian operator, on the dual on-shell Bethe vector is given by
\begin{equation}
\label{eq:eigenvalue-bethe}
\bra{\Psi(\bm{v})}\mathcal{T}(z) = \Lambda(z; \bm{v}) \bra{\Psi(\bm{v})} \; , 
\end{equation}
where \(\Lambda(z; \bm{v}) \) is the transfer matrix eigenvalue.

We construct \(n+1\) functions obtained from products between the
on-shell and each off-shell Bethe states
\begin{equation}
  \bm{\zeta}_j(\bm{u}_j, \bm{v}) 
  = \bracket{\Psi(\bm{v})}{\Psi(\bm{u}_j)}\; ,\qquad j =1, \dots, n+1\; .
\end{equation}
These partially on-shell scalar products are called \emph{Slavnov
products}.  Henceforth, the functional dependence of these functions
will be omitted, that is \(\bm{\zeta}_j\equiv \bm{\zeta}_j(\bm{u}_j,
\bm{v})\).

In this work we only consider integrable models where the action of
the transfer matrix on a generic off-shell bethe states can be
expanded as
\begin{subequations}
\begin{equation}
\label{eq:scalar-exp}
  \mathcal{T}(u_j)\ket{\Psi(\bm{u}_j)}  = \sum_{k=1}^{n+1} L_{jk} \ket{\Psi(\bm{u}_k)} \; ,
\end{equation}
where \(L_{jk}\) are coefficients, and the off-diagonal elements
\(L_{jk}\) with \(j\neq k\) contain the unwanted
terms~\cite{Slavnov:2019aba}. From the Algebraic Bethe Ansatz
construction, the non-diagonal terms must vanish when the Bethe
equations are satisfied, and this is basically the definition of the
algebraic bethe equations. Consequently, only the diagonal
coefficients \(L_{jj}\) survive, and we conclude that these terms must
be equal to the eigenvalue of the tranfer matrix.  Therefore
\begin{equation}
  \mathcal{T}(u_j)\ket{\Psi(\bm{u}_j)} =  \Lambda(u_j; \bm{u}_j) \ket{\Psi(\bm{v})} + 
  \sum_{\substack{k=1\\ k\neq j}}^{n+1} L_{jk} \ket{\Psi(\bm{u}_k)} \; . 
\end{equation}
\end{subequations}

This class of models includes many of the most familiar spin chains
with rational and trigonometric \(R\)-matrices -- including those
models discussed in the introduction. One important class of models
that does not fall into this classification is defined by
elliptic \(R\)-matrices, although some of these models still have slavnov
products with determinantal representations.

\begin{proposition}
The Slavnov products \(\bm{\zeta}_j\) satisfy a system of linear
equations~\cite{Slavnov:2019aba}.
\end{proposition}
\paragraph{\textbf{\emph{Proof}}} Let us first use that the
transfer matrix can act on the on-shell bra \(\bra{\Psi(\bm{v})}\)
on on the off-shell ket \(\ket{\Psi(\bm{u}_j)}\), then  
\begin{equation}
\label{eq:scalar-prod}
  \bra{\Psi(\bm{v})} \left(\mathcal{T}(u_j)\ket{\Psi(\bm{u}_j)} \right) =
  \left(\bra{\Psi(\bm{v})}\mathcal{T}(u_j)\right)\ket{\Psi(\bm{u}_j)}
  \qquad j =1, \dots, n+1\; .
\end{equation}
The right-hand side of this equation can be simplified with the eigenvalue
expression~(\ref{eq:eigenvalue-bethe}), that is
\begin{equation}
  \left(\bra{\Psi(\bm{v})}\mathcal{T}(u_j)\right)\ket{\Psi(\bm{u}_j)}  =
\Lambda(u_j; \bm{v}) \bracket{\Psi(\bm{v})}{\Psi(\bm{u}_j)} = 
\Lambda(u_j; \bm{v}) \bm{\zeta}_j\; . 
\end{equation}

Let us now use the expansion~(\ref{eq:scalar-exp}) on the left-hand
side of~(\ref{eq:scalar-prod}). Therefore, we have
\begin{equation}
  \bra{\Psi(\bm{v})} \left(\mathcal{T}(u_j)\ket{\Psi(\bm{u}_j)} \right) =
  \sum_{k=1}^{n+1} L_{jk} \bm{\zeta}_k \; .
\end{equation}

Puttting all these facts together, we can write the
expression~(\ref{eq:scalar-prod}) as
\begin{equation}
  \sum_{k=1}^{n+1} L_{jk} \bm{\zeta}_k  = 
  \Lambda(u_j; \bar{v}) \bm{\zeta}_j\quad \Rightarrow \quad
  \sum_{k=1}^{n+1} M_{jk} \bm{\zeta}_k = 0 \; ,
\end{equation}
where we have defined the coefficients
\begin{equation}
  M_{jk} = L_{jk} - \delta_{jk}\Lambda(u_j; \bm{v})\; . 
\end{equation}
In simple terms, this expression shows that the Slavnov products
satisfy a linear system.  We can write this system in a matrix form as
\begin{equation}
\label{eq:linear-system}
\bm{M} \bm{\zeta} = 
  \left(\bm{L} - \bm{\Lambda} \right) \bm{\zeta} = \bm{0}\; ,
\end{equation}
where
\begin{subequations}
  \begin{equation}
    \bm{L} = 
 \begin{pmatrix}
   L_{1,1} & \cdots & L_{1,n+1}\\
   L_{2,1} & \cdots & L_{2,n+1}\\
   \vdots & \vdots & \vdots\\
   L_{n+1, 1} & \cdots & L_{n+1, n+1}\\
 \end{pmatrix}  \quad , \quad
    \bm{\Lambda} = \mathrm{diag}(\Lambda_1, \Lambda_2, \dots, \Lambda_{n+1})
\end{equation}
with \(\Lambda_j \equiv \Lambda(u_j; \bm{v})\), and
\begin{equation}
  \bm{\zeta} = (\bm{\zeta}_1, \dots , \bm{\zeta}_{n+1})^T\qquad \, 
  \bm{0} = (0, \dots , 0)^T\; .
\end{equation}
\end{subequations}
This completes the proof of their proposition. 
 \qed

There are two important consequences for us now. First of all, if we
assume that this system has a nontrivial solution, it will be
expressed in terms of minors of the matrix \(\bm{M}\).

Moreover, since the system is homogeneous, the solutions are
determined up to multiplicative factors, which we can fix by
requiring that the final result yields a KP tau function. In the
original works of Foda and collaborators, e.g.,~\cite{Foda:2009zz,
  Wheeler:2010vmq}, the authors perform a series of redefinitions to
achieve the same goal.

Evidently, being a determinant is not enough to guarantee that the
Slavnov product is a tau function.  In order to establish this
result, we need to specify the models. Following the original
work~\cite{Belliard:2019bfz}, we will restrict the current analysis to rational models
-- we will see that the calculations are overwhelming even in this
simple case. The generalization to open boundary conditions and for
trigonometric models is straightforward, see~\cite{Slavnov:2019aba} for more
details.

\subsection{Rational models}

Let us start with an R-matrix of the form
\begin{equation}
  R(u,v) = \mathbb{1} + g(u,v) \mathcal{P}\; , \qquad
  g(u,v) = \frac{c}{u - v}
\end{equation}
where \(c\) is a constant, \(\mathbb{1}\) is the identity and
\(\mathcal{P}\) is the permutation operator~\cite{Slavnov:2019aba}.
The eigenvalues of the transfer matrix have the form
\begin{equation}
  \Lambda(z, \bm{v}) = g(z, \bm{v}) \mathcal{Y}(z; \bm{v})\; ,
\end{equation}
where the function \(\mathcal{Y}(z; \bm{v})\) is symmetric over the
Bethe roots \(\bm{v}\), and has a linear dependence on each Bethe root
\(v_j\).

Generically, we write the \(\mathcal{Y}\)-functions as 
\begin{equation}
\label{eq:yfunc-exp}
  \mathcal{Y}(z; \bm{v}) = \sum_{p=0}^n \alpha_p(z) \sigma_p^{(n)}(\bm{v})\; , 
\end{equation}
where \(\sigma_p^{(n)}(\bm{v})\) are elementary symmetric polynomials
in \(\bm{v}\), and \(\alpha_p(z)\) are free functional parameters.
The most important point for us is that the functions \(\alpha_p(z)\), 
and consequently the \(\mathcal{Y}\)-function, are regular in the limit
\(z \to v_j \in \bm{v}\). The XXX spin chain has been discussed
in~\cite{Belliard:2019bfz}, and the authors show that \(\alpha_p(z)\)
are polynomials of degree \(2n\).

Let us also define the product 
\begin{equation}
  g(z, \bm{v}) = \prod_{v_i\in \bm{v}} g(z; v_i)\; . 
\end{equation}
Now, it is easy to see that each Bethe root \( v_i \in \bm{v}\) is a
pole of \(g(z;\bm{v})\). Therefore, if \(\Gamma\) is a contour
containing all the Bethe roots \(\{v_j\}_{j=1}^n\), we have
\begin{equation}
  \frac{1}{2 \pi i}\oint_{\Gamma} \mathrm{d}z g(z; \bm{v}) = \sum_{j=1}^n
  \prod_{v_i \in \bm{v}\setminus \{v_j\}} g(z, v_i)\; .
\end{equation}
Consequently, the coefficients \(L_{jk}\) are given by
\begin{equation}
  L_{jk} = g(u_k, \bm{u}_k) \mathcal{Y}(u_k; \bm{u}_j)\; ,
\end{equation}
and from this expression one can calculate the matrix \(\bm{M}\).

It has also been shown~\cite{Belliard:2019bfz} that for \(\det(\bm{M})
= 0\) and \(\mathrm{rank}(\bm{M}) = n\), the Slavnov products are
given by
\begin{equation}
 \bm{\zeta}_\ell = \phi(\bm{v}) \tilde{\Delta}(\bm{u}_\ell)\hat{\Omega}_\ell\; , \qquad \ell = 1, \dots, n+1\; .
\end{equation}
Let us now explain the different terms in this expression.  First of
all, \(\phi(\bm{v})\) is a function of the Bethe roots and its
particular form is not important in out analysis. Moreover, we have
\begin{equation}
  \tilde{\Delta}(\bm{u}_\ell) = \prod_{\substack{u_j, u_j \in \bm{u}_\ell \\ j>k}} \frac{c}{u_j - u_k}\; .
\end{equation}
We can also absorb the product of \(c\) into a new constant
\(c_0\), and it is easy to see that \(\tilde{\Delta}(\bm{u}_\ell)/c_0\) is the
inverse of the Vandermonde determinant \(\Delta(\bm{u}_\ell) =
\prod_{j>k}(u_j - u_k)\).

We now define the \(n\times (n+1)\) matrix \(\bm{\Omega}\) by specifying its components
\begin{equation}
  \label{eq:omega-matrix-def}
  \Omega_{jk}(u_k; \bm{v}) = g(u_k, v_j) \mathcal{Y}(u_k; \{u_k, \bm{v}_j\})\; ,
  \quad j= 1, \dots, n\; , \quad k =1, \dots, n+1\; .
\end{equation}
The matrices \(\hat{\Omega}_\ell\) are minors of
\(\bm{\Omega}\) with the \(\ell\)-th column excluded, in other words,
\begin{equation}
 \hat{\Omega}_\ell = \det_{k\neq \ell} \Omega_{jk}\; . 
\end{equation}
All in all, we use these determinants to define \(n+1\)
\emph{normalized Slavnov products} as follows
\begin{equation}
  \label{eq:slavnov-tau}
  \bm{\tau}^{(\ell)}(\bm{u}_\ell; \bm{v}) \equiv \frac{1}{c_0}\frac{\bm{\zeta}_\ell}{\phi(\bm{v})}
  = \frac{\hat{\Omega}_{\ell}}{\Delta(\bm{u}_\ell)}\; , \quad \ell = 1, \dots, n+1 \; .
\end{equation}
This is the most important result for our discussion; and our goal now
is to show that each \(\bm{\tau}^{(\ell)}(\bm{u}_\ell; \bm{v})\) is a
tau-function of the KP-hierarchy.

\section{Tau functions}

In order to proceed, let us simplify the notation and organize the
results described above. First of all, observe that we can define
\(n\) functions \(\Omega_j(z)\) of the form
\begin{equation}
\begin{split}
  \Omega_{j}(z) \equiv
  \Omega_{j}(z; \bm{v}) & = g(z, v_j) \mathcal{Y}(z; \{z, \bm{v}_j\})\\
   & = \frac{c}{z - v_j} \mathcal{Y}(z; \{z, \bm{v}_j\})\; , \qquad j =1, \dots, n \; .
\end{split}
\end{equation}
Let us also denote \(\mathcal{Y}(z; \{z, \bm{v}_j\}) \equiv
\mathcal{Y}_j(z)\), and note that we have \(n\) functions
\(\mathcal{Y}_j(z)\) whose defining property is the absence of the
\(j\)-th Bethe root \(v_j\).  It is also convenient to consider
\(c=1\), consequently \(c_0=1\).

Moreover, the expansion of the
\(\mathcal{Y}\)-functions~(\ref{eq:yfunc-exp}) yields
\begin{equation}
  \mathcal{Y}(z; \{z,\bm{v}_j\}) = \sum_{p=0}^n \alpha_p(z) \sigma_p^{(n)}(\{z,\bm{v}_j\})\; . 
\end{equation}
Using that the elementary symmetric polynomials satisfies the
relations
\begin{equation}
  \sigma_p^{(n)}(\{z,\bm{v}_j\}) = \sigma_p^{(n-1)}(\bm{v}_j) + z \sigma_{p-1}^{(n-1)}(\bm{v}_j) \; . 
\end{equation}
we write
\begin{equation}
  \label{eq:y-functions-alpha}
  \mathcal{Y}(z; \{z,\bm{v}_j\})  = 
\sum_{p=0}^{n} \alpha_p(z)
  \left(\sigma_p^{(n-1)}(\bm{v}_j) + z \sigma_{p-1}^{(n-1)}(\bm{v}_j) \right) \; .
\end{equation}
The elementary symmetric polynomials also satisfy
\(\sigma_n^{(n-1)}(\bm{v}_j)=0\) and \(\sigma_{-1}^{(n)}(\bm{v}_j)=
0\), and it yields
\begin{equation}
  \label{eq:y-functions}
  \mathcal{Y}_j(z)  = 
  \sum_{p=0}^{n-1} \beta_p(z) \sigma_p^{(n-1)}(\bm{v}_j) \; , \qquad 
\beta_p(z) = \alpha_p(z) + z \alpha_{p+1}(z) \; .
\end{equation}

Putting all these facts together, we have
\begin{equation}
\begin{split}
  \Omega_{j}(z; \bm{v}) & = \frac{\mathcal{Y}_j(z) }{z - v_j} \\ 
& = \frac{1}{z - v_j} \sum_{p=0}^{n-1} \beta_p(z) \sigma_p^{(n-1)}(\bm{v}_j) \; .
\end{split}
\end{equation}
The factor \((z - v_j)^{-1}\) introduces the dependence on the Bethe root \(v_j\).  
Moreover, one can observe that
\begin{equation}
\Res_{z=v_k} \left( \Omega_j(z)\right) =  \delta_{jk}\mathcal{Y}_j(v_j)\; .
\end{equation}

\subsection{Alternant determinant expression for the Slavnov products}

Let us now write the matrix \(\bm{\Omega}\) defined
in~(\ref{eq:omega-matrix-def}) as
\begin{equation}
  \bm{\Omega} =
  \begin{pmatrix}
    \Omega_1(u_1) & \dots & \Omega_1(u_\ell) & \dots  & \Omega_1(u_n) & \Omega_1(u_{n+1}) \\
    \Omega_2(u_1) & \dots & \Omega_2(u_\ell) & \dots  & \Omega_2(u_n) & \Omega_2(u_{n+1}) \\
          \vdots &       &   \vdots    &    &     \vdots       & \vdots   \\
    \Omega_n(u_1) & \dots & \Omega_n(u_\ell)& \dots  & \Omega_n(u_n) & \Omega_n(u_{n+1})
  \end{pmatrix}\; .
\end{equation}
From this expression, we can also define \(n+1\) square matrices
\(\bm\Omega^{(\ell)}(\bm{u}_\ell) \equiv \bm\Omega^{(\ell)}(\bm{u}_\ell; \bm{v})\), for
\(\ell = 1, \dots, n+1\), by deleting the \(\ell\)-th column of
\(\bm{\Omega}\). That is
\begin{equation}
  \bm{\Omega}^{(\ell)} =
  \begin{pmatrix}
    \Omega_1(u_1) & \dots & \Omega_1(u_{\ell-1}) & \Omega_1(u_{\ell+1})&\dots & \Omega_1(u_{n+1}) \\
    \vdots &  & \vdots & \vdots & & \vdots \\
    \Omega_n(u_1) & \dots & \Omega_n(u_{\ell-1}) & \Omega_n(u_{\ell+1})&\dots & \Omega_n(u_{n+1})
  \end{pmatrix}\; .
\end{equation}
Henceforth, we will only refer to the matrix
\(\bm{\Omega}\) defined by the components
\begin{equation}
  \label{eq:omega-matrix}
  \Omega_{jk} = \frac{\mathcal{Y}_j(z_k)}{z_k - v_j}\; .
\end{equation}

Finally, we can write~(\ref{eq:slavnov-tau}) as 
\begin{equation}
\label{eq:slavnov-tau-functions}
  \bm{\tau}^{(\ell)}(\bm{u}_\ell; \bm{v}) = \frac{\det [\Omega_j((\bm{u}_\ell)_k)]|_{j,k=1}^n }{\Delta(\bm{u}_\ell)} \; ,
\end{equation}
where \((\bm{u}_\ell)_k \) is the \(k\)-th component of the vector
\(\bm{u}_\ell\).  More explicitly, remember that \(\bm{u} = (u_1,
\dots, u_{n+1})\) and that \(\bm{u}_\ell = \bm{u}_\ell\setminus
\{u_\ell\} = (u_1, \dots, u_{\ell-1}, u_{\ell+ 1}, \dots, u_{n+1})\),
we have
\begin{equation}
  (\bm{u}_\ell)_k =
  \left\{ 
  \begin{matrix}
    u_k & \text{if} & k < \ell\\
    u_{k+1} & \text{if} & k > \ell
  \end{matrix}
  \right. \; .
\end{equation}
It is convenient to write \(\bm{z}^{(\ell)} = (z^{(\ell)}_1,
z^{(\ell)}_2, \dots, z^{(\ell)}_n) = \bm{u}_{\ell}\). Moreover, it is
worth noting that the normalized Slavnov product
\(\bm{\tau}^{(\ell)}\) is completely independent of the parameter
\(u_\ell\). On the other hand, these functions are not independent from each
other, since \(\bm{z}^{(\ell)} \cap \bm{z}^{(\ell')} = \bm{u} \setminus
\{ u_\ell, u_{\ell'}\} \).

Consequently, we write the normalized Slavnov products as 
\begin{equation}
\begin{split}
\label{eq:slav-tau-function}
  \bm{\tau}^{(\ell)}(\bm{z}^{(\ell)}, \bm{v})
  & = \frac{1}{\Delta(\bm{z^{(\ell)}})} \det[\Omega_j(z^{(\ell)}_k) ]_{j,k=1}^n \\
  & = \frac{1}{\Delta(\bm{z}^{(\ell)})} \det\left[\frac{\mathcal{Y}_j(z^{(\ell)}_k)}{z^{(\ell)}_k - v_j} \right]_{j,k=1}^n
  \; \quad \ell = 1, \dots, n+1 \; .
\end{split}
\end{equation}
This is the first result of out work. 

All the normalized Slavnov products~(\ref{eq:slav-tau-function}) take
the form of \emph{alternant determinants} of the functions
\(\{\Omega_j\}_{j=1}^n\) divided by Vandermonde determinants.  This
result establishes that the normalized slavnov
products~(\ref{eq:slav-tau-function}) are tau functions of the KP
hierarchy. Indeed, it is well documented that, given a set of generic
functions \(\{\phi_j\}_{j=1}^n\), expressions of the type
\begin{equation}
\tau(\bm{w}) =
\frac{\det_{i,j} \phi_i(w_j)}{\Delta(\bm{w})}\; ,
\end{equation}
where \(\{w_j\}_{j=1}^n\) are complex parameters, satisfy the Hirota
bilinear equation. Consequently, they are tau functions of the KP
hierarchy.  This proposition has been extensively discussed in the
literature, for example~\cite{Segal:1985aga, Kharchev:1991cy,
  Araujo:2021ghu} and references therein.

In~\cite{Segal:1985aga, Kharchev:1991cy, Alexandrov:2014cwa}, the
authors explored a family of functions \(\{ \phi_i\}_{i=1}^n\) that
parametrize points in a Grassmannian space. These functions also play
an important role in the definition of the Baker-Akhiezer
functions. However, the asymptotic behavior of the functions
\(\Omega_j\), which are relevant to our discussion, indicates that
they do not belong to the class considered in the references above.
Therefore, we must investigate their properties and the corresponding
Baker-Akhiezer functions specific to our discussion.  We begin this
analysis below.

Given that each function \(\bm{\tau}^{(\ell)}(\bm{z}^{(\ell)}, \bm{v})\), for
\(\ell = 1, \dots, n+1\), is a distinct but related tau function, one
can define a vector
\begin{equation}
  \bm{T}(\bm{Z}, \bm{v}) = 
\begin{bmatrix}
\bm{\tau}^{(1)}(\bm{z}^{(1)}, \bm{v}) \\ 
\bm{\tau}^{(2)}(\bm{z}^{(2)}, \bm{v}) \\ 
\vdots \\ 
\bm{\tau}^{(n+1)}(\bm{z}^{(n+1)}, \bm{v})
\end{bmatrix}
\end{equation}
where \(\bm{Z} = (\bm{z}^{(1)}, \bm{z}^{(2)}, \dots,
\bm{z}^{(n+1)})\). This is a vector of tau functions of the KP
hierarchy.

\begin{conjecture}
This observation suggests that there is, in fact, an
underlying multicomponent KP hierarchy in this context, and that the
linear system~(\ref{eq:linear-system}), which we can write as
\begin{equation}
\left( \bm{L} - \bm{\Lambda} \right) \bm{T} = \bm{0},
\end{equation}
defines a reduction of this multicomponent KP hierarchy. This
reduction ultimately describes constraints that relate the different
sets of parameters \(\bm{z}^{(\ell)}\).  We have not explored this
idea in this work, but we hope to return to this problem in future
publications.
\end{conjecture}

\subsection{Tau functions expansion of the Slavnov product}

Let us now fix a component \(\ell\) and omit this index in
equation~(\ref{eq:slav-tau-function}). We also assume that the Bethe
roots are non-degenerate; that is, for \(j \neq k\), we necessarily
have \(v_j \neq v_k\). In other words, each Bether root \(v_j\)
corresponds to a simple poles of the Slavnov product.

Fix a particular coefficient of \(\bm{z}\), say \(z_l\), and consider a Laplace
expansion of~(\ref{eq:slav-tau-function}) along the \(l\)-th column. That is
\begin{equation}
\label{eq:minor-expansion}
\begin{split}
  \bm{\tau}(\bm{z}, \bm{v})
  & = \frac{1}{\Delta(\bm{z})} \det[\bm{\Omega}(\bm{z}) ] \\
  & = \frac{1}{\Delta(\bm{z})} \sum_{j=1}^n (-1)^{j + l} \frac{\mathcal{Y}_j(z_l)}{z_l - v_j  } 
  \det[\hat{\Omega}_{j, l}]\; ,
\end{split}
\end{equation}
where \(\det[\hat{\Omega}_{j,l}]\) denotes the \((j, l)\)-minor of
\(\bm{\Omega}\). We can now extract the residue of the tau function
with respect to \(z_l\) at the point \(v_j\); that is
\begin{equation}
  \Res_{z_l=v_j}(\Delta(\bm{z})\tau(\bm{z}, \bm{v})) = (-1)^{j+l} \mathcal{Y}_j(v_j) \det[\hat{\Omega}_{j, l}]\;. 
\end{equation}

\begin{subequations}
Additionally, decompose the Vandermonde determinant as
\begin{equation}
  \Delta(\bm{z}) = \prod_{j>k}(z_j - z_k) = \Delta(\bm{z}_l) \prod_{r<l}(z_l - z_r) \prod_{s>l}(z_s - z_l)\; ,
\end{equation}
where \(\bm{z}_l = \bm{z}\setminus \{z_l\} = (z_1, \dots, z_{l-1},
z_{l+1}, \dots, z_n) \), and we also define the function
\begin{equation}
  \Xi_l(z_l,\bm{z}_l) = \prod_{r<l}(z_l - z_r) \prod_{r>l}(z_r - z_l)\; , 
\end{equation}
then 
\begin{equation}
  \Delta(\bm{z}) = \Xi_l(z_l, \bm{z}_l) \Delta(\bm{z}_l) \; .
\end{equation}
\end{subequations}

Inserting this decomposition into~(\ref{eq:minor-expansion}), we find
\begin{equation}
\label{eq:minor-expansion-2}
  \bm{\tau}(\bm{z}, \bm{v})
   = \frac{1}{\Xi_l(z_l, \bm{z}_l)} \sum_{j=1}^n (-1)^{j + l} \frac{\mathcal{Y}_j(z_l)}{z_l - v_j} 
\left(\frac{1}{\Delta(\bm{z}_l)} \det[\hat{\Omega}_{j, l}]\right)\; .
\end{equation}

By defining the \(n\) sets of functions
\begin{equation}
  \begin{split}
    \hat{\bm{\Omega}}_{(j)}(z) & = \left(\Omega_1(z), \dots, \Omega_{j-1}(z), \Omega_{j+1}(z), \dots, \Omega_n(z) \right)\; ,
    \quad j =1, \dots, n\; , 
  \end{split}
\end{equation}
we can further simplify the minor
expansion~(\ref{eq:minor-expansion-2}); that is
\begin{equation}
  \label{eq:slavnov-tauexp}
\begin{split}
  \bm{\tau}(\bm{z}, \bm{v})
  & = \frac{1}{\Xi_l(z_l, \bm{z}_l)} \sum_{j=1}^n (-1)^{j + l} \frac{\mathcal{Y}_j(z_l)}{z_l - v_j} 
    \left(\frac{1}{\Delta(\bm{z}_l)} \det[\hat{\bm{\Omega}}_{(j) r}(\bm{z}_{(l) s} ]_{r, s=1}^n\right)\\ 
  & = \frac{1}{\Xi_l(z_l, \bm{z}_l)} \sum_{j=1}^n (-1)^{j + l} \frac{\mathcal{Y}_j(z_l)}{z_l - v_j} 
    \tilde{\bm{\tau}}_j(\bm{z}_l, \bm{v})\; ,
\end{split}
\end{equation}
where \(\hat{\bm{\Omega}}_{(j)r}\) is the \(r\)-th component of
\(\hat{\bm{\Omega}}_{(j)}\) and \(\hat{z}_{(l)s}\) is the \(s\)-th
component of \(\bm{z}_{(l)}\). Moreover, in the second line we have
defined the object
\begin{equation}
  \label{eq:basis-tau}
    \tilde{\bm{\tau}}_j(\bm{z}_l, \bm{v}) = 
\frac{1}{\Delta(\bm{z}_l)} \det[\hat{\bm{\Omega}}_{(j)}(\bm{z}_{(l)})]\; .
\end{equation}
Finally, we have
\begin{equation}
  \Res_{z_l = v_j}(\tau(\bm{z}, \bm{v}))
  = \frac{(-1)^{j + l} \mathcal{Y}_j(v_j)}{\Xi_l(v_j, \bm{z}_l)}
  \tilde{\bm{\tau}}_j(\bm{z}_l, \bm{v})\; .
\end{equation}

We can also organize the parameters \(\bm{z}\) and consider these
points close to the corresponding Bethe roots \(\bm{v}\). Therefore 
\begin{subequations}
\begin{equation}
  \Res_{z_j = v_j}(\tau(\bm{z}, \bm{v}))
  = \frac{\mathcal{Y}_j(v_j)}{\Xi_j(v_j, \bm{z}_j)}
  \tilde{\bm{\tau}}_j(\bm{z}_j, \bm{v})\; ,
\end{equation}
or yet 
\begin{equation}
  \tilde{\bm{\tau}}_j(\bm{z}_j, \bm{v})
  = \frac{\Xi_j(v_j, \bm{z}_j)}{\mathcal{Y}_j(v_j)}
  \Res_{z_j = v_j}(\tau(\bm{z}, \bm{v}))\; .
\end{equation}
\end{subequations}

Expression~(\ref{eq:slavnov-tauexp}) is one of the main results in
this work.  It is easy to see that each term
\(\tilde{\bm{\tau}}_j(\bm{z}_l, \bm{v})\) is a tau function itself,
and serves as a basis for the Slavnov's product. Additionally, these
basis tau functions \(\tilde{\bm{\tau}}_j(\bm{z}_l, \bm{v})\) are, by
construction, completely independent of \(z_l\). This result
essentially shows that, given a \(\mathcal{Y}\)-function (which is
related to the eigenvalue of the transfer matrix), one can construct a
basis of tau functions that span the Slavnov products in the
corresponding integrable system. Moreover, this expansion ensures that
the resulting Slavnov product is also a tau function of the KP
hierarchy.

\section{Homogeneous limit: Wronskian formula}

In the previous section, we discussed the limit in which the
parameters \(\bm{z}\) approach the Bethe roots \(\bm{v}\).  We now
consider a different limit, where all variables in the set \(\bm{z}\)
tend to a single variable -- that is, \(z_k\to z_1\) for \(k=2,\dots,
n\). The analysis follows the ideas of~\cite{Izergin:1991rs}. 

Let us first consider the case \(z_2 \to z_1\). We perform a series
expansion around \(z_1\), in which the second column
of~(\ref{eq:slav-tau-function}) becomes
\begin{subequations}
\begin{equation}
  \bm{\tau}(\bm{z}, \bm{v}) = \frac{1}{\Delta(\bm{z})}
  \det
  \begin{pmatrix}
    \frac{\mathcal{Y}_1(z_1)}{z_1 - v_1}  & \frac{\mathcal{Y}_1(z_1)}{z_1 - v_1} + (z_2 - z_1) \frac{\mathcal{Y}_1^{(1)}(z_1)}{z_1 - v_1}
    + \mathcal{O}(\delta^2) &
    \frac{\mathcal{Y}_1(z_3)}{z_3 - v_1} & \dots & \frac{\mathcal{Y}_1(z_n)}{z_n - v_1}\\
    \frac{\mathcal{Y}_2(z_1)}{z_1 - v_2} & \frac{\mathcal{Y}_2(z_1)}{z_1 - v_2} + (z_2 - z_1) \frac{\mathcal{Y}_2^{(1)}(z_1)}{z_1 - v_2} 
    + \mathcal{O}(\delta^2) &
    \frac{\mathcal{Y}_2(z_3)}{z_3 - v_2} &  \dots & \frac{\mathcal{Y}_2(z_n)}{z_n - v_2}\\
    &  \vdots & & \\
    \frac{\mathcal{Y}_n(z_1)}{z_1 - v_n} & \frac{\mathcal{Y}_n(z_1)}{z_1 - v_n} + (z_2 - z_1) \frac{\mathcal{Y}_n^{(1)}(z_1)}{z_1 - v_n} 
    + \mathcal{O}(\delta^2) &
    \frac{\mathcal{Y}_n(z_3)}{z_3 - v_n} &  \dots & \frac{\mathcal{Y}_n(z_n)}{z_n - v_n}
  \end{pmatrix}\; ,
\end{equation}
where we have written \(z_2 - z_1 = \delta \to 0\). Moreover, let us
denote by \(\mathcal{Y}^{(n)}(z)\) the \(n\)-th derivative of
\(\mathcal{Y}(z)\) with respect its argument.

One can immediately see that the second column is equal to the first
column plus terms proportional to the factor
\(\delta = z_2 - z_1\). Using elementary column operations, we find
\begin{equation}
  \bm{\tau}_h(\bm{z}, \bm{v}) = \lim_{z_2 \to z_1}\frac{(z_2 - z_1)}{\Delta(\bm{z})}
  \det
  \begin{pmatrix}
    \frac{\mathcal{Y}_1(z_1)}{z_1 - v_1}  & \frac{\mathcal{Y}_1^{(1)}(z_1)}{z_1 - v_1} &
    \frac{\mathcal{Y}_1(z_3)}{z_3 - v_1} & \dots & \frac{\mathcal{Y}_1(z_n)}{z_n - v_1}\\
    \frac{\mathcal{Y}_2(z_1)}{z_1 - v_2} & \frac{\mathcal{Y}_2^{(1)}(z_1)}{z_1 - v_2} &
    \frac{\mathcal{Y}_2(z_3)}{z_3 - v_2} &  \dots & \frac{\mathcal{Y}_2(z_n)}{z_n - v_2}\\
    &  \vdots & & \\
    \frac{\mathcal{Y}_n(z_1)}{z_1 - v_n} & \frac{\mathcal{Y}_n^{(1)}(z_1)}{z_1 - v_n} &
    \frac{\mathcal{Y}_n(z_3)}{z_3 - v_n} &  \dots & \frac{\mathcal{Y}_n(z_n)}{z_n - v_n}
  \end{pmatrix}\; .
\end{equation}
We can now repeat the same reasoning for \(z_3\to z_1=z_2\). We find
that the third column becomes a linear combination of the first and
second columns, along with terms involving derivatives multiplied by
the factor \(\delta^2 = (z_3 - z_2)(z_3 - z_1)\). All in all, we have
\begin{equation}
  \bm{\tau}_h(\bm{z}, \bm{v}) = \lim_{z_2, z_3 \to z_1}\frac{(z_2 - z_1)(z_3 - z_1)(z_3 - z_2)}{\Delta(\bm{z})}
  \det
  \begin{pmatrix}
    \frac{\mathcal{Y}_1(z_1)}{z_1 - v_1}  & \frac{\mathcal{Y}_1^{(1)}(z_1)}{z_1 - v_1} &
    \frac{\mathcal{Y}^{(2)}_1(z_1)}{z_1 - v_1} & \dots & \frac{\mathcal{Y}_1(z_n)}{z_n - v_1}\\
    \frac{\mathcal{Y}_2(z_1)}{z_1 - v_2} & \frac{\mathcal{Y}_2^{(1)}(z_1)}{z_1 - v_2} &
    \frac{\mathcal{Y}_2^{(2)}(z_1)}{z_1 - v_2} &  \dots & \frac{\mathcal{Y}_2(z_n)}{z_n - v_2}\\
    &  \vdots & & \\
    \frac{\mathcal{Y}_n(z_1)}{z_1 - v_n} & \frac{\mathcal{Y}_n^{(1)}(z_1)}{z_1 - v_n} &
    \frac{\mathcal{Y}_n^{(2)}(z_1)}{z_1 - v_n} &  \dots & \frac{\mathcal{Y}_n(z_n)}{z_n - v_n}
  \end{pmatrix}\; .
\end{equation}

By applying this procedure iteratively to each column, we find that
the multiplicative factors cancel the Vandermonde determinant, and the
homogeneous limit becomes
\begin{equation}
  \bm{\tau}_h(z_1, \bm{v}) =
  \det
  \begin{pmatrix}
    \frac{\mathcal{Y}_1(z_1)}{z_1 - v_1}  & \frac{\mathcal{Y}_1^{(1)}(z_1)}{z_1 - v_1} &
    \frac{\mathcal{Y}^{(2)}_1(z_1)}{z_1 - v_1} & \dots & \frac{\mathcal{Y}_1^{(n-1)}(z_1)}{z_1 - v_1}\\
    \frac{\mathcal{Y}_2(z_1)}{z_1 - v_2} & \frac{\mathcal{Y}_2^{(1)}(z_1)}{z_1 - v_2} &
    \frac{\mathcal{Y}_2^{(2)}(z_1)}{z_1 - v_2} &  \dots & \frac{\mathcal{Y}_2^{(n-1)}(z_1)}{z_1 - v_2}\\
    &  \vdots & & \\
    \frac{\mathcal{Y}_n(z_1)}{z_1 - v_n} & \frac{\mathcal{Y}_n^{(1)}(z_1)}{z_1 - v_n} &
    \frac{\mathcal{Y}_n^{(2)}(z_1)}{z_1 - v_n} &  \dots & \frac{\mathcal{Y}^{(n-1)}_n(z_1)}{z_1 - v_n}
  \end{pmatrix}\; .
\end{equation}
Finally, we write \(z_1 \equiv w\) and using elementary row operations
we find
\begin{equation}
  \bm{\tau}_h(w, \bm{v}) =
  \frac{1}{\prod_{k=1}^n(w - v_k)}
  \mathcal{W}[\mathcal{Y}_1, \mathcal{Y}_2, \dots, \mathcal{Y}_n](w)
  \; .
\end{equation}
\end{subequations}
An interesting aspect of this expression is that its poles coincide
precisely with the Bethe roots \(\bm{v}\). Moreover, the homogeneous
limit can be readily constructed from the Algebraic Bethe Ansatz,
since it depends only on the transfer matrix
eigenvalues. Additionally, the functions
\(\{\mathcal{Y}_j(w)\}_{j=1}^n\) are linearly independent if and only
if the Bethe roots are non-degenerate.

\section{Examples}

Let us now consider the cases \(n = 2\) and \(n = 3\) to gain further
insight into the problem.

\subsection{Case n=2}
In this case, we have \(\mathcal{Y}_j\) for \(j = 1, 2\) and two Bethe
roots \(\bm{v} = (v_1, v_2)\). Moreover, we define the two sets
\(\bm{v}_1 = \{v_2\}\) and \(\bm{v}_2 = \{v_1\}\). From
equation~(\ref{eq:y-functions}), we have
\begin{subequations}
\begin{equation}
  \begin{split}
    \mathcal{Y}_j(z) =
    \alpha_0(z) \sigma_0^{(1)}(\bm{v}_j) +  
    \alpha_1(z)( \sigma_1^{(1)}(\bm{v}_j) + z \sigma_0^{(1)}(\bm{v}_j)) +
    \alpha_2(z) z \sigma_1^{(1)}(\bm{v}_j) \; . 
  \end{split}
\end{equation}
Furthermore, the explicit formulas for the elementary symmetric
polynomials are \(\sigma_0^{(1)}(x) = 1\) and \(\sigma_1^{(1)}(x) =
x\). Therefore
\begin{equation}
  \begin{split}
    \mathcal{Y}_1(z) & = \alpha_0(z) + z \alpha_1(z)  + v_2 \left( \alpha_1(z) + z \alpha_2(z) \right) \\ 
    \mathcal{Y}_2(z) & = \alpha_0(z) + z \alpha_1(z)  + v_1 \left( \alpha_1(z) + z \alpha_2(z) \right) \; .
  \end{split}
\end{equation}
Consequently, the normalized Slavnov product becomes
\begin{equation}
\begin{split}
  \bm{\tau}(z_1, z_2; v_1 , v_2)
  & = \frac{1}{z_2 - z_1} \left(\frac{\mathcal{Y}_1(z_1) \mathcal{Y}_2(z_2)}{(z_1 - v_1)(z_2 - v_2)} - 
  \frac{\mathcal{Y}_1(z_2) \mathcal{Y}_2(z_1)}{(z_1 - v_2)(z_2 - v_1)} \right)\\
  & = \frac{1}{z_2 - z_1}
  \left(
  \frac{\mathcal{Y}_1(z_1)}{(z_1 - v_1)} \widetilde{\bm{\tau}}_1(z_2, v_1, v_2) - 
  \frac{\mathcal{Y}_2(z_1)}{(z_1 - v_2)} \widetilde{\bm{\tau}}_2(z_2, v_1, v_2) \right)\; ,
\end{split}
\end{equation}
where 
\begin{equation}
 \widetilde{\bm{\tau}}_1(z_2, v_1, v_2) = \frac{\mathcal{Y}_2(z_2)}{(z_2 - v_2)} \quad \textrm{and}  \qquad 
 \widetilde{\bm{\tau}}_2(z_2, v_1, v_2) = \frac{\mathcal{Y}_1(z_2)}{(z_2 - v_1)}  \; .
\end{equation}
\end{subequations}

From this expression, it is easy to extract the residues of \(z_1\) at
one of the Bethe roots. It is also straightforward to see that the
homogeneous limit, that is, \(z_2 \to z_1 \equiv w\), yields
\begin{equation}
  \bm{\tau}(w; v_1 , v_2) =
  \frac{1}{(w- v_1)(w- v_2)}\mathcal{W}[\mathcal{Y}_1, \mathcal{Y}_2](w)\; . 
\end{equation}
It is also elementary to see that the functions \(\{\mathcal{Y}_1,
\mathcal{Y}_2\}\) are linearly independent as long as \(v_2 \neq
v_1\), which is guaranteed by the assumption that the Bethe roots are
distinct.

\subsection{Case n=3}
Now we have \(\mathcal{Y}_j\) for \(j=1,2,3\), and three Bethe roots
\(\bm{v} = (v_1, v_2, v_3)\). Moreover, we define the sets \(\bm{v}_1
= (v_2, v_3)\), \(\bm{v}_2 = (v_1, v_3)\), and \(\bm{v}_3 = (v_1,
v_2)\). Then, it is easy to see that
\begin{equation}
\begin{split}
  \mathcal{Y}_j(z)
  & = (\alpha_0(z) + z \alpha_1(z))\sigma_0^{(2)}(\bm{v}_j)   + (\alpha_1(z) + z \alpha_2(z))\sigma_1^{(2)}(\bm{v}_j)  + \\
  & \quad + (\alpha_2(z) + z \alpha_3(z))\sigma_2^{(2)}(\bm{v}_j) \; ,
\end{split}
\end{equation}
with
\begin{equation}
  \sigma_0^{(2)}(x, y) = 1\; ,\qquad 
  \sigma_1^{(2)}(x, y) = x + y\; ,\qquad 
  \sigma_2^{(2)}(x, y) = x^2 + xy  + y^2\; . 
\end{equation}

Hence
\begin{subequations}
\begin{equation}
\begin{split}
\bm{\tau} & (z_1, z_2, z_3 ; v_1 , v_2, v_3)  =
\frac{1}{(z_2 - z_1)(z_3 - z_1)(z_3 - z_2)}\times \\
& \times \left[
    \frac{\mathcal{Y}_1(z_1)}{z_1 - v_1}
    \det\begin{bmatrix}
    \frac{\mathcal{Y}_2(z_2)}{(z_2 - v_2)} & \frac{\mathcal{Y}_2(z_3)}{(z_3 - v_2)}\\
    \frac{\mathcal{Y}_3(z_2)}{(z_2 - v_3)} & \frac{\mathcal{Y}_3(z_3)}{(z_3 - v_3)}
    \end{bmatrix}
    -  
    \frac{\mathcal{Y}_2(z_1)}{z_1 - v_2}
    \det\begin{bmatrix}
    \frac{\mathcal{Y}_1(z_2)}{(z_2 - v_1)} & \frac{\mathcal{Y}_1(z_3)}{(z_3 - v_1)}\\
    \frac{\mathcal{Y}_3(z_2)}{(z_2 - v_3)} & \frac{\mathcal{Y}_3(z_3)}{(z_3 - v_3)}
    \end{bmatrix} \right.\\
    & \qquad + \left.
    \frac{\mathcal{Y}_3(z_1)}{z_1 - v_2}
    \det\begin{bmatrix}
    \frac{\mathcal{Y}_1(z_2)}{(z_2 - v_1)} & \frac{\mathcal{Y}_1(z_3)}{(z_3 - v_1)}\\
    \frac{\mathcal{Y}_2(z_2)}{(z_2 - v_2)} & \frac{\mathcal{Y}_2(z_3)}{(z_3 - v_2)}
    \end{bmatrix} 
\right] \; .
\end{split}
\end{equation}
We reorganize this expression as follows:
\begin{equation}
\begin{split}
\bm{\tau} (z_1, z_2, z_3 ; v_1 , v_2, v_3)  & =
\frac{1}{(z_2 - z_1)(z_3 - z_1)}
\left[
    \frac{\mathcal{Y}_1(z_1)}{z_1 - v_1}
    \left(
    \frac{1}{(z_3 - z_2)}
    \det\begin{bmatrix}
    \frac{\mathcal{Y}_2(z_2)}{(z_2 - v_2)} & \frac{\mathcal{Y}_2(z_3)}{(z_3 - v_2)}\\
    \frac{\mathcal{Y}_3(z_2)}{(z_2 - v_3)} & \frac{\mathcal{Y}_3(z_3)}{(z_3 - v_3)}
    \end{bmatrix}
    \right) - \right. \\
    & - 
    \frac{\mathcal{Y}_2(z_1)}{z_1 - v_2}
    \left(
    \frac{1}{(z_3 - z_2)}
    \det\begin{bmatrix}
    \frac{\mathcal{Y}_1(z_2)}{(z_2 - v_1)} & \frac{\mathcal{Y}_1(z_3)}{(z_3 - v_1)}\\
    \frac{\mathcal{Y}_3(z_2)}{(z_2 - v_3)} & \frac{\mathcal{Y}_3(z_3)}{(z_3 - v_3)}
    \end{bmatrix} \right) \\
    & +
    \left.
    \frac{\mathcal{Y}_3(z_1)}{z_1 - v_2}
    \left(
    \frac{1}{(z_3 - z_2)}
    \det\begin{bmatrix}
    \frac{\mathcal{Y}_1(z_2)}{(z_2 - v_1)} & \frac{\mathcal{Y}_1(z_3)}{(z_3 - v_1)}\\
    \frac{\mathcal{Y}_2(z_2)}{(z_2 - v_2)} & \frac{\mathcal{Y}_2(z_3)}{(z_3 - v_2)}
    \end{bmatrix} 
    \right)
\right]\; .
\end{split}
\end{equation}
\end{subequations}

Finally, we conclude that the basis tau functions are
\begin{subequations}
  \begin{equation}
	\tilde{\bm{\tau}}_1(\bm{z}_1) = \frac{1}{(z_3 - z_2)} 
        \det\begin{bmatrix}
        \frac{\mathcal{Y}_2(z_2)}{(z_2 - v_2)} & \frac{\mathcal{Y}_2(z_3)}{(z_3 - v_2)}\\
        \frac{\mathcal{Y}_3(z_2)}{(z_2 - v_3)} & \frac{\mathcal{Y}_3(z_3)}{(z_3 - v_3)}
        \end{bmatrix} 
  \end{equation}
  \begin{equation}
	\tilde{\bm{\tau}}_2(\bm{z}_1) = \frac{1}{(z_3 - z_2)} 
        \det\begin{bmatrix}
        \frac{\mathcal{Y}_1(z_2)}{(z_2 - v_1)} & \frac{\mathcal{Y}_1(z_3)}{(z_3 - v_1)}\\
        \frac{\mathcal{Y}_3(z_2)}{(z_2 - v_3)} & \frac{\mathcal{Y}_3(z_3)}{(z_3 - v_3)}
       \end{bmatrix}
  \end{equation}
  \begin{equation}
	\tilde{\bm{\tau}}_3(\bm{z}_1) = \frac{1}{(z_3 - z_2)} 
        \det\begin{bmatrix}
        \frac{\mathcal{Y}_1(z_2)}{(z_2 - v_1)} & \frac{\mathcal{Y}_1(z_3)}{(z_3 - v_1)}\\
        \frac{\mathcal{Y}_2(z_2)}{(z_2 - v_2)} & \frac{\mathcal{Y}_2(z_3)}{(z_3 - v_2)}
        \end{bmatrix} \; . 
  \end{equation}
\end{subequations}

With these expressions, we can consider the homogeneous limit \(z_1,
z_2, z_3 \to w\). Consider first the case \(z_3 \to z_2 = w\), then we
know that
\begin{equation}
  \begin{split}
  \tilde{\bm{\tau}}_1(w; v_1 , v_2, v_3) & = \frac{1}{(w - v_2)(w - v_3)}\mathcal{W}[\mathcal{Y}_2, \mathcal{Y}_3](w)\\
  \tilde{\bm{\tau}}_2(w; v_1 , v_2, v_3) & = \frac{1}{(w - v_2)(w - v_3)}\mathcal{W}[\mathcal{Y}_1, \mathcal{Y}_3](w)\\
  \tilde{\bm{\tau}}_3(w; v_1 , v_2, v_3) & = \frac{1}{(w - v_1)(w - v_2)}\mathcal{W}[\mathcal{Y}_1, \mathcal{Y}_2](w)
  \; .
  \end{split}
\end{equation}
It is immediate to see that the basis tau functions correspond to
Slavnov products for the case \(n=2\). We also take the limit \(z_1\to
w\); then
\begin{equation}
  \bm{\tau}(w; v_1 , v_2, v_3) = 
  \frac{1}{(w - v_1)(w - v_2)(w - v_3)} \mathcal{W}[\mathcal{Y}_1, \mathcal{Y}_2, \mathcal{Y}_3](w)\; .
\end{equation}

Of course, we could continue the calculations for other cases, but it
is now clear that this explicit analysis becomes cumbersome quite
quickly. This also explains why we restrict our analysis to the
rational cases.

\section{Baker-Akhiezer function}

This section addresses some properties of the Baker-Akhiezer functions
associated with the tau functions derived above. While many aspects of
these functions deserve thorough examination, here we focus on their
most essential properties.

\subsection{Integral representation of the tau functions}

We now aim to express the Slavnov products defined above in terms of
the following coordinates
\begin{equation}
  t_p = \frac{1}{p}\sum_{j=1}^n z_j^p\; ,
\end{equation}
the so-called Miwa coordinates. Let us also define the function
\begin{equation}
  \xi(\bm{t}, \lambda) = \sum_{p=1}^\infty t_p \lambda^p\; , 
\end{equation}
where \(\lambda\) is a complex parameter.

Therefore
\begin{equation}
  \begin{split}
    e^{\xi(\bm{t}, \lambda)} & = \exp\left( \sum_{p=1}^\infty t_p \lambda^p \right)
    = \exp\left( \sum_{p=1}^\infty \sum_{j=1}^n \frac{1}{p} z^p \lambda^p \right) \\
    & = \exp\left( \sum_{j=1}^n \sum_{p=1}^\infty  \frac{1}{p} z_j^p \lambda^p \right)
    = \exp\left( - \sum_{j=1}^n \ln( 1 -  z_j \lambda ) \right) \\
    & =  \prod_{j=1}^n \frac{1}{1 -  z_j \lambda} \; .
  \end{split}
\end{equation}
It is easy to see that these functions have simple poles at \( \lambda = z_j^{-1}\).

\begin{proposition}
\label{prop:integral-rep}
It has been established that tau functions in the alternant
form~(\ref{eq:slav-tau-function}) admits the following integral
representation
\begin{subequations}
\begin{equation}
  \label{eq:tau-function-miwa}
  \bm{\tau}(\bm{t}, \bm{v}) =
    \det_{j,k}\left(
    \oint_{\gamma_k} \frac{d w}{2\pi i} e^{\xi(\bm{t}, w^{-1})} \frac{w^{-j}\mathcal{Y}_k(w)}{w - v_k} \right)\; ,
\end{equation}
where 
\begin{equation}
  \xi(\bm{t}, w^{-1}) = \sum_{p=1}^\infty t_p w^{-p} \quad \Leftrightarrow \quad
  \xi(\bm{z}, w^{-1}) = w^n \prod_{j=1}^n (w - z_j)^{-1} \; . 
\end{equation}
In the above integral, we consider that integration curve \(\gamma_k\)
encloses all poles except the Bethe root \(v_k\).
\end{subequations}
\end{proposition}

\paragraph{\textbf{\emph{Proof}}}

As far as we know, this result was first established
in~\cite{ZinnJustin:2002cr, ZinnJustin:2002pk}; see
also~\cite{Zinnjustin2009, Araujo:2021ghu} for further
discussions. Here, we present a proof of this result for pedagogical
reasons and to address some differences that arise due to our
conventions.

In terms of the \(z\)-coordinates, the integral~(\ref{eq:tau-function-miwa}) becomes
\begin{equation}
  \label{eq:tau-function}
  \bm{\tau}(\bm{z}, \bm{v}) =
    \det_{j,k}\left(
    \oint_{\gamma_k} \frac{d w}{2\pi i}
    \prod_{l=1}^n \frac{1}{w - z_l} \frac{w^{n-j}\mathcal{Y}_k(w)}{w - v_k} \right) \; . 
\end{equation}
Since \(n \geq j\), the point \(w = 0\) is not a pole of the
integrand. All in all, the integration contour \(\gamma_k\) encloses
the points \(\{z_j\}_{j=1}^n\).

Let us define the matrix\footnote{
The choice of the transposition is an aesthetic one. We want 
to derive a result expressed as a matrix product with the components \(\Omega_{kl}\) on the left.}
\(\bm{\mathcal{K}}^T\) by its components as follows:
\begin{equation}
\label{eq:k-matrix}
    \mathcal{K}_{kj} = 
\oint_{\gamma_k} \frac{d w}{2\pi i} e^{\xi(\bm{z}, w^{-1})} \frac{w^{-j}\mathcal{Y}_k(w)}{w - v_k}\; .
\end{equation}
We can now carry out the integration
\begin{equation}
  \begin{split}
    \mathcal{K}_{kj}
    = & \oint_{\gamma_k} \frac{d w}{2\pi i} \frac{w^{n-j}}{\prod_{s} (w - z_s)}
    \frac{\mathcal{Y}_k(w)}{w - v_k} = \sum_{\ell =1}^n \frac{z_l^{n-j}}{\prod_{s\neq \ell} (z_\ell - z_s)}
    \frac{\mathcal{Y}_k(z_\ell)}{z_\ell - v_k}\\
    & = \sum_{\ell = 1}^n  \Omega_{k \ell} \left( \frac{z_\ell^{n-j}}{\prod_{s\neq \ell} (z_\ell - z_s)} \right)\; ,
  \end{split}
\end{equation}
where in the second line we have used the \(\bm{\Omega}\)-matrix
defined in~(\ref{eq:omega-matrix}).

We can now see that the matrix \(\bm{\mathcal{K}}\) can be understood
as the product of two other matrices. Consequently, we have
\begin{equation}
  \begin{split}
  \det \bm{\mathcal{K}} & = \det(\bm{\Omega}) \det_{l,j}\left( \frac{z_l^{n-j}}{\prod_{j\neq l} (z_l - z_j)}\right) \\
  & = \det(\bm{\Omega}) \det_{l, j}(z_l^{n-j}) \prod_{\substack{j, l =1 \\ j \neq l}}^n (z_l - z_j)^{-1} \; .
  \end{split}
\end{equation}
We also have the identities 
\begin{subequations}
  \begin{equation}
	\det_{l,j}(z_l^{n-j}) = (-1)^{n(n-1)/2} \Delta(\bm{z})\; ,
  \end{equation}
  and
  \begin{equation}
	\prod_{\substack{j, l = 1 \\ j \neq l}}^n (z_l - z_j) = (-1)^{n(n-1)/2} \Delta(\bm{z})^2\; . 
  \end{equation}
\end{subequations}
Combining all these expressions, we finally see that
\begin{equation}
  \det \bm{\mathcal{K}} = \frac{\det \bm{\Omega}}{\Delta(\bm{z})}\; ,
\end{equation}
then \(\bm{\tau}(\bm{z}, \bm{v}) = \det\bm{\mathcal{K}}\), that is
preciselly the expression~(\ref{eq:slav-tau-function}) for a fixed
\(\ell\). \\ \qed

\begin{lemma}
\label{lemma:k-invert}
The matrix \(\bm{\mathcal{K}}\) defined via~(\ref{eq:k-matrix}) is
invertible. This follows immediately from the relation
\(\bm{\tau}(\bm{z}, \bm{v}) = \det \bm{\mathcal{K}}\).
\end{lemma}

One advantage of the integral
representation~(\ref{eq:tau-function-miwa}) is that it makes it easier
to consider the limit of the infinite chain, \(n\to \infty\).

\subsection{Baker-Akhiezer  in \(z\)-coordinates}

The Baker-Akhiezer (BA) function is defined through the
Japanese formula~\cite{Babelon:2003qtg, Harnad:2021tau, Zabrodin2018}.
\begin{equation}
\label{eq:ba-function}
  \psi(\bm{t},\bm{v}; \lambda) = e^{\xi(\bm{t}, \lambda)}
  \frac{\bm{\tau}(\bm{t} - [\lambda^{-1}], \bm{v})}{\bm{\tau}(\bm{t}, \bm{v})}\; ,
\end{equation}
where 
\begin{equation}
  \bm{t} - [\lambda^{-1}] = 
  \{t_1 - \lambda^{-1}, t_2 - \lambda^{-2}/ 2 ,  t_2 - \lambda^{-3}/3, \dots,  t_p - \lambda^{-p}/p, \dots \}\; .
\end{equation}
We can write these components as
\begin{equation}
  t_p - \frac{\lambda^{-p}}{p} = \frac{1}{p} \sum_{j=1}^n z_j^ p - \frac{\lambda^{-p}}{p} \; , 
\end{equation}
therefore
\begin{equation}
  \begin{split}
    e^{\xi(\bm{t} - [\lambda^{-1}], w^{-1})} & = \exp\left[ \sum_{p=1}^\infty \left(t_p  - \frac{\lambda^{-p}}{p}\right) w^{-p} \right]\\
    & = e^{\xi(\bm{t}, w^{-1})} \exp\left[ - \sum_{p=1}^\infty \frac{\lambda^{-p}}{p} w^{-p} \right] \\
    & = e^{\xi(\bm{t}, w^{-1})} \exp\left[ \ln ( 1 - 1 / (\lambda w)) \right]
      = \left( 1 - \frac{1}{\lambda w}\right)  \prod_{j=1}^n \frac{1}{1 -  z_j /w} \\
    & = \frac{w^{n-1}}{\lambda } \left( \lambda w - 1 \right)  \prod_{j=1}^n \frac{1}{w -  z_j} \; .
  \end{split}
\end{equation}

Finally, we express the shifted tau functions
\begin{subequations}
\begin{equation}
  \bm{\tau}(\bm{t} - [\lambda^{-1}], \bm{v}) =
  \det_{j,k}\left[
  \oint_{\gamma_k} \frac{dw}{2\pi i} e^{\xi(\bm{t}, w^{-1})} \frac{w^{- j -1} ( \lambda w - 1) \mathcal{Y}_k(w)}{\lambda(w - v_k)} 
  \right]
\end{equation}
in terms of the \(z\)-coordinates as
\begin{equation}
\begin{split}
  \bm{\tau}(\bm{z}, \bm{v}; \lambda^{-1})
  & = \det_{j,k}\left[\oint_{\gamma_k} \frac{dw}{2\pi i} 
    \frac{( \lambda w - 1) w^{n-j-1}\mathcal{Y}_k(w)}{\lambda(w - v_k)} \prod_{s=1}^n \frac{1}{w - z_s} \right]\\
  & = \det_{j,k}\left[
    \oint_{\gamma_k} \frac{dw}{2\pi i}  
    \frac{w^{n-j-1}\widehat{\mathcal{Y}}_k(w; \lambda)}{(w - v_k)} \prod_{s=1}^n \frac{1}{w - z_s} 
    \right]\; ,
\end{split}
\end{equation}
\end{subequations}
where we have defined the functions
\begin{equation}
  \widehat{\mathcal{Y}}_k(w; \lambda)
  = \frac{( \lambda w - 1)}{\lambda} \mathcal{Y}_k(w) \; , \qquad k =1, \dots, n\;. 
\end{equation}
From these expressions, it is easy to see that for \(j = n\), the
point \(w = 0\) is a pole, and we must include it in the integration
contour \(\gamma_k\).

Define the matrix \(\widetilde{\bm{\mathcal{K}}}^T\) by its components
\begin{equation}
\label{eq:k-tilde}
  \begin{split}
    \widetilde{\mathcal{K}}_{kj} = &
\oint_{\gamma_k} \frac{dw}{2\pi i} e^{\xi(\bm{t}, w^{-1})} \frac{w^{- j -1} ( \lambda w - 1) \mathcal{Y}_k(w)}{\lambda(w - v_k)}
    \\
    = & \oint_{\gamma_k} \frac{dw}{2\pi i}  
    \frac{( \lambda w - 1) w^{n-j-1}\mathcal{Y}_k(w)}{\lambda(w - v_k)} \prod_{s=1}^n \frac{1}{w - z_s}  \\
    = &
    \oint_{\tilde \gamma_k} \frac{dw}{2\pi i}
    \frac{w^{n-j-1}\widehat{\mathcal{Y}}_k(w; \lambda)}{(w - v_k)} \prod_{s=1}^n \frac{1}{w - z_s} 
    + \delta_{nj} F_k(\lambda) \\
    = & \sum_{\ell = 1}^n  \widehat{\Omega}_{k \ell} \left( \frac{z_\ell^{n-j}}{\prod_{s\neq \ell} (z_\ell - z_s)} \right)
    + F_k(\lambda)\delta_{nj}  \\
  \end{split}
\end{equation}
where \(\tilde{\gamma}_k\) is a deformed contour that does not enclose
the point \(w = 0\), and
\begin{equation}
\label{eq:omehat-comp}
  \widehat{\Omega}_{k\ell}
  = \frac{1}{z_\ell} \frac{\widehat{\mathcal{Y}}_k(z_\ell, \lambda)}{z_\ell - v_k}
  = \frac{(\lambda z_\ell -1)}{\lambda z_\ell} {\Omega}_{k\ell} 
\; , \qquad 
 F_k(\lambda) = (-1)^{n+1} \frac{\widehat{\mathcal{Y}}_k(0; \lambda)}{v_k }  \prod_{s=1}^n z_s^{-1} \; .
\end{equation}
Moreover, it is also convenient to define the matrix
\(\widehat{\bm{\Omega}}\) formed by the components
\(\widehat{\Omega}_{ij}\) and
the vector \(\bm{F} = (F_1, \dots , F_n)\), 
with the components defined above in~(\ref{eq:omehat-comp}).

Additionally, let us define yet another matrix,
\(\widehat{\bm{\mathcal{K}}}^T\), by
\begin{equation}
    \widehat{\mathcal{K}}_{kj} = 
    \sum_{\ell = 1}^n  \widehat{\Omega}_{k \ell}
    \left( \frac{z_\ell^{n-j}}{\prod_{s\neq \ell} (z_\ell - z_s)} \right) \; .
\end{equation}

We conclude that \(\widetilde{\bm{\mathcal{K}}}\) can be derived from
\(\widehat{\bm{\mathcal{K}}}\) by adding, component-wise, the vector
\(\bm{F} = (F_1, \dots , F_n)\) to the \(n\)-th row.  We write this
operation as
\begin{equation}
  \widetilde{\bm{\mathcal{K}}} = \widehat{\bm{\mathcal{K}}} + \bm{0}_{[ \bm{F} \to n\text{-row}]}\; . 
\end{equation}
Here we use the notation \(\mathcal{\bm{A}}_{[ \bm{F} \to j\text{-row}]}\) 
to denote the replacement of the \(j\)-th row of the matrix 
\(\mathcal{\bm{A}}\) with the vector \(\bm{F}\). In the expression above, 
we have considered this operation with the null matrix \(\bm{0}\).
By the multilinearity of the determinant, we finally have
\begin{equation}
  \det \widetilde{\bm{\mathcal{K}}} = 
    \det \widehat{\bm{\mathcal{K}}} + 
    \det \widehat{\bm{\mathcal{K}}}_{[\bm{F} \to n\text{-row}]}\; .
\end{equation}

It is now easy to see that \(\det \widehat{\bm{\mathcal{K}}}\) is
proportional to the Slavnov product, that is,
\begin{equation}
    \det \widehat{\bm{\mathcal{K}}}  = 
    \det(\widehat{\bm{\Omega}})
    \det_{l,j}
    \left( \frac{z_l^{n-j}}{\prod_{s\neq l} (z_l - z_s)} \right)\\
    = \frac{\det(\widehat{\bm{\Omega}})}{\Delta(\bm{z})}\;. 
\end{equation}
The above determinant is itself a tau function, but we can also write
it as
\begin{equation}
    \det \widehat{\mathcal{K}} = 
    \frac{\det(\Omega)}{\Delta(\bm{z})}
    \prod_{l=1}^n \left( 1 - \frac{1}{\lambda z_l} \right) 
    = \bm{\tau}(\bm{z}, \bm{v}) \prod_{l=1}^n \left( 1 - \frac{1}{\lambda z_l} \right) \; . 
\end{equation}
Furthermore, we can simplify this expression using the elementary
symmetric polynomials \(\sigma_p^{(n)}\); that is,
\begin{equation}
  \prod_{l=1}^n \left( 1 - \frac{1}{\lambda z_l} \right) = 1 + 
  \sum_{p=1}^n (-1)^p \lambda^{-p} \sigma_p^{(n)}(\bm{z}^{-1})\; , 
\end{equation}
where \( \bm{z}^{-1} = \{z_1^{-1}, z_2^{-1}, \dots , z_n^{-1}\}\). 

Collecting all these facts, we finally write the Baker-Akhiezer
function as
\begin{equation}
  \psi(\lambda; \bm{z}, \bm{v}) =
  e^{\xi(\bm{t}, \lambda)}  \left[ 1 + 
  \sum_{p=1}^n (-1)^p \lambda^{-p} \sigma_p^{(n)}(\bm{z}^{-1})
    + \frac{1}{\bm{\tau}(\bm{z}, \bm{v})}
    \det \widehat{\bm{\mathcal{K}}}_{[\bm{F} \to n\text{-row}]}
  \right]\; .
\end{equation}
From this expression we can see that all the poles are located at the
point \(\lambda = 0\).

\subsection{Baker-Akhiezer in \(t\)-coordinates}

In this section, we find a better expression for the Baker-Akhiezer
function. Our main goal now is to reconsider the above calculations
using the Miwa coordinates.

Let us begin our exploration with the definitions~(\ref{eq:k-matrix})
and~(\ref{eq:k-tilde}). Let us write
\begin{equation}
\label{eq:ktilde-split}
  \widetilde{\mathcal{K}}_{kj} = \mathcal{K}_{kj}
  - \frac{1}{\lambda} \oint_{\gamma_k} \frac{dw}{2\pi i}
    e^{\xi(\bm{t}, w^{- 1})} \frac{w^{-j -1 }\mathcal{Y}_k(w)}{w - v_k} \; .
\end{equation}
Therefore, let us define one more matrix,
\(\check{\bm{\mathcal{K}}}^T\), with components given by
\begin{equation}
  \check{\mathcal{K}}_{jk} =
  - \oint_{\gamma_k} \frac{dw}{2\pi i}
    e^{\xi(\bm{t}, w^{- 1})} \frac{w^{-j -1 }\mathcal{Y}_k(w)}{w - v_k} \; .
\end{equation}
Therefore,~(\ref{eq:ktilde-split}) becomes
\begin{equation}
  \widetilde{\mathcal{K}}_{kj} = 
  \mathcal{K}_{kj}  + \frac{1}{\lambda}
  \check{\mathcal{K}}_{kj} \; , 
\end{equation}
that yields
\begin{equation}
  \widetilde{\bm{\mathcal{K}}} = 
  \bm{\mathcal{K}} + \frac{1}{\lambda}
  \check{\bm{\mathcal{K}}}\; .
\end{equation}
Moreover, by Lemma~\ref{lemma:k-invert}, \(\bm{\mathcal{K}}\) is
invertible; therefore,
\begin{equation}
  \widetilde{\bm{\mathcal{K}}} = 
  \bm{\mathcal{K}}\left( \mathbb{1} + 
  \frac{1}{\lambda}\bm{\mathcal{K}}^{-1}\check{\bm{\mathcal{K}}} \right)\; .
\end{equation}

Combining these results, we find that the shifted tau-functions are
\begin{equation}
  \bm{\tau}(\bm{t} - [\lambda^{-1}], \bm{v}) =
  \bm{\tau}(\bm{t}, \bm{v})
  \det \left( \mathbb{1} + \frac{1}{\lambda}
  \bm{\mathcal{K}}^{-1} \check{\bm{\mathcal{K}}} \right)\; .
\end{equation}
We finally conclude that the Bakher-Akhiezer
function~(\ref{eq:ba-function}) can be written as
\begin{equation}
\label{eq:ba-fredholm}
  \psi(\bm{t},\bm{v}; \lambda) = e^{\xi(\bm{t}, \lambda)}
  \det \left( \mathbb{1} + 
    \frac{1}{\lambda} \bm{\mathcal{M}}
  \right)\; , 
\end{equation}
where \(\bm{\mathcal{M}} =
\bm{\mathcal{K}}^{-1}\check{\bm{\mathcal{K}}}\) is a
finite-dimensional matrix depending on the coordinates
\(\bm{t}\).

Let us now use the elementary property
\begin{subequations}
\begin{equation}
  \begin{split}
    \det \left( \mathbb{1} + \frac{1}{\lambda} \bm{\mathcal{M}}\right)
    & = \exp \left[ \Tr \ln \left( \mathbb{1} + \frac{1}{\lambda} \bm{\mathcal{M}} \right) \right]
    = \exp\left[ \sum_{l=1}^n \ln \left( 1 + \frac{\mu_l}{\lambda} \right)\right]\\
    & = \prod_{l=1}^n \left( 1 + \frac{\mu_l}{\lambda} \right) \; ,
  \end{split}
\end{equation}
\end{subequations}
where \(\bm{\mu} = \{\mu_l\}_{l=1}^n\) are the eigenvalues of the
matrix \(\bm{\mathcal{M}}\), and these obviously depend on the Bethe
roots \(\bm{v}\) and parameters \(\bm{t}\).  Therefore, we have
\begin{equation}
  \psi(\bm{t},\bm{v}; \lambda) = e^{\xi(\bm{t}, \lambda)}
  \left(1 + \sum_{k=1}^n \frac{\xi_k(\bm{t})}{\lambda^k} 
  \right)\; \qquad \xi_k(\bm{t}) = \sigma_k^{(n)}(\bm{\mu})\; .
\end{equation}

Alternatively, we can also express the determinant as
\begin{equation}
  \begin{split}
    \det \left( \mathbb{1} + \frac{1}{\lambda} \bm{\mathcal{M}}\right)
    & = \exp\left[ \sum_{l=1}^n \ln \left( 1 + \frac{\mu_l}{\lambda} \right)\right]
    = \exp\left[ \sum_{l=1}^n \sum_{k=1}^\infty \frac{(-1)^{k-1}}{k} \left(\frac{\mu_l}{\lambda} \right)^k\right] \\ 
    & = \exp\left[ \sum_{k=1}^\infty \frac{(-1)^{k-1}}{\lambda^k} \gamma_k\right] \; ,
  \end{split}
\end{equation}
where we define the coordinates as 
\begin{equation}
\gamma_k = \frac{1}{k} \sum_{l=1}^n \mu_l^k\; .
\end{equation}

Then, we write the Baker-Akhiezer as 
\begin{equation}
  \psi(\bm{t},\bm{v}; \lambda) = e^{\xi(\bm{t}, \lambda)}
  \left(1 + \sum_{k\geq 1} \frac{\sigma_k(\bm{\gamma})}{\lambda^k} 
  \right)\; .
\end{equation}
Notice that this second expression does not explicitly depend on the
size \(n\) of the Slavnov product. Consequently, the functions
\(\xi_k(\bm{t})\) can be expressed in terms of the coordinates
\(\bm{\gamma}\), where the parameter \(n\) is now implicitly contained
in their definition.

An advantage of this formulation is that it provides a more suitable
framework to consider the thermodynamic limit \(n \to \infty\) with
\(n/L \to 0\). In fact, expression~(\ref{eq:ba-fredholm}) suggests
that in this limit the Baker-Akhiezer function can be represented as a
Fredholm determinant. We are currently investigating this question and
hope to report new results soon.

\section{Discussion}

In this work, we have discussed several properties of the Slavnov
products arising in quantum integrable models and their deep
connection with the tau functions of the KP hierarchy. We can
summarize our main findings as follows. Our initial result
demonstrates that the general structure of these tau functions can be
expressed in terms of an alternant matrix. This general formulation
firmly establishes the identification of the Slavnov product as a tau
function within the framework of the KP classical integrable hierarchy.

We have also proved that these Slavnov products admit a basis
expansion in terms of other tau functions, thereby establishing that
we are dealing with a particularly distinguished object within this
framework. Moreover, we have discussed a conjecture suggesting the
existence of a multicomponent KP hierarchy underlying all of our
results.

The behaviour of these functions near the Bethe roots of the quantum
integrable systems, as well as the homogeneous limit of the Slavnov
product, has also been discussed. Finally, we conclude our work with a
brief analysis of the Baker-Akhiezer function. The most important
result for us is that we have shown that the Baker-Akhiezer function,
modulo a universal multiplicative factor, also admits a determinantal
form.

Evidently, there are many results that can be extended in this work,
and several aspects that deserve further investigation. Let us list
some of these problems, ranging from relatively straightforward
applications to more substantial challenges.

A simple problem to be discussed is the physical meaning of the
solutions of the KP equation that can be constructed using the tau
functions explored in this work. Perhaps one might consider both
analytical and numerical approaches, since the calculation of the
determinants becomes complicated even for relatively small values of
\(n\).

Another important aspect of this work is to achieve a better
understanding of the Baker-Akhiezer functions associated with the
Slavnov products. Here, we have only scratched the surface of these
objects, and a full description of their properties is still
lacking. We are currently investigating the thermodynamic limit of
this system. In particular, we are studying the Slavnov product and
its interpretation from the viewpoint of the Baker-Akhiezer
function. For example, how to properly describe the limit \(L \to
\infty\) and \(N \to \infty\), with \(N\) growing sufficiently
slower than \(L\). I believe that the matrix \(\mathcal{M}\) is a
trace-class operator, allowing us to express the Baker-Akhiezer
function as a Fredholm determinant. We hope to report new results
on this line of investigation soon.

Finally, there is a more challenging problem to be addressed: the
description of the elliptic case. While some determinantal formulas
are known for the partially on-shell scalar product of Bethe states,
it is not yet clear whether these objects are also related to tau
functions of integrable hierarchies.  This characterization is another
problem we are currently investigating, and we hope to have some
results to report in the future.

\subsubsection*{Acknowledgments}

I would like to express my sincere gratitude to the Department of
Physics at Fluminense Federal University for providing an excellent
research environment and for their support of this work.

\bibliographystyle{utphys}
\bibliography{bib-database.bib}

\end{document}